\documentclass[dvipsnames,letterpaper, 10 pt, conference]{ieeeconf}

\IEEEoverridecommandlockouts
\usepackage[table]{xcolor}

\usepackage{cite}
\usepackage{amsmath,amssymb,amsfonts}
\usepackage{algorithmic}
\usepackage{textcomp}
\usepackage{graphicx}
\usepackage{mathrsfs}
\usepackage{tikz}
\usepackage[vlined,ruled]{algorithm2e}
\usepackage{subfigure}
\usepackage{url}
\usepackage{dsfont}
\usepackage{bbm}
\usepackage{booktabs}
\usepackage{array}
\usepackage{yfonts}
\usepackage[normalem]{ulem}
\usepackage{arydshln,leftidx,mathtools}
\usepackage{multicol}
\usepackage{amsmath}
\usepackage{stfloats}
\usepackage{bbm}
\usepackage{comment}
\usepackage{subfigure}
\usepackage{manyfoot}
\usepackage{listings}
\usepackage{float}
\usepackage[most]{tcolorbox}
\usepackage{placeins}
\usetikzlibrary{positioning,arrows.meta,shapes.misc}
\usepackage{pifont}

\definecolor{codegreen}{rgb}{0,0.6,0}
\definecolor{codegray}{rgb}{0.5,0.5,0.5}
\definecolor{codepurple}{rgb}{0.58,0,0.82}
\definecolor{backcolour}{rgb}{0.95,0.95,0.92}
\definecolor{ourshade}{gray}{1.92}

\lstdefinestyle{mystyle}{
    backgroundcolor=\color{backcolour},
    commentstyle=\color{codegreen},
    keywordstyle=\color{magenta},
    numberstyle=\tiny\color{codegray},
    stringstyle=\color{codepurple},
    basicstyle=\ttfamily\footnotesize,
    breakatwhitespace=false,
    breaklines=true,
    captionpos=b,
    keepspaces=true,
    numbers=left,
    numbersep=5pt,
    showspaces=false,
    showstringspaces=false,
    showtabs=false,
    tabsize=2
}

\definecolor{myBlue}{RGB}{49,130,189}

\DontPrintSemicolon
\SetKw{KwContinue}{continue}
\SetKw{KwAnd}{and}
\SetKw{KwRet}{return}
\SetKwInput{KwProc}{Procedure}

\usepackage{tabularx}
\usepackage{multirow}
\usepackage{makecell}

\newcommand\aamsout{\bgroup\markoverwith{\textcolor{violet}{\rule[0.5ex]{2pt}{1pt}}}\ULon}

\DeclareFontFamily{U}{stix2bb}{}
\DeclareFontShape{U}{stix2bb}{m}{n} {<-> stix2-mathbb}{}

\DeclareSymbolFont{bbold}{U}{bbold}{m}{n}
\DeclareSymbolFontAlphabet{\mathbbold}{bbold}

\newcommand\oprocendsymbol{\hbox{$\square$}}
\newcommand\oprocend{\relax\ifmmode\else\unskip\hfill\fi\oprocendsymbol}

\tcbset{
  panel/.style n args={2}{
    enhanced,
    arc=10pt,
    boxrule=0.5pt,
    colframe=#1!80!black,
    colback=#1!12,
    colbacktitle=#1!25,
    coltitle=black,
    fonttitle=\bfseries\normalsize,
,
    title={#2},
    title filled,
    titlerule=1.2pt,
    toptitle=2mm, bottomtitle=2mm,
    left=10pt,right=10pt,top=10pt,bottom=10pt,
    before skip=8pt, after skip=10pt,
  }
}

\newtcolorbox{TaskBox}[1]{panel={blue}{#1}}
\newtcolorbox{EnvBox}[1]{panel={green}{#1}}

\newtcolorbox{StageBox}{
  enhanced,
  colback=white,
  colframe=black!60,
  boxrule=0.8pt,
  arc=6pt,
  left=8pt,right=8pt,top=6pt,bottom=6pt,
  before skip=6pt, after skip=8pt,
}

\newcommand*{\QEDA}{\hfill\ensuremath{\blacksquare}}%

\DeclareNewFootnote{R}[roman]

\graphicspath{{figs/}}

\makeatletter
\let\NAT@parse\undefined
\makeatother
\usepackage[colorlinks,urlcolor=blue, linkcolor=blue, citecolor=blue]{hyperref}

\begin{document}
\title{\LARGE \bf Spectrogram-Based Joint Detection, Localization, and Classification of Events in Continuously Recorded IBR Waveforms\vspace{-0.15cm}}

\author{\spaceskip=0.9\fontdimen2\font\relax Shivanshu~Tripathi,~\IEEEmembership{Graduate Student Member,~IEEE,} Maziar~Raissi, and~Hamed~Mohsenian-Rad,~\IEEEmembership{Fellow,~IEEE}
\thanks{S.~Tripathi and H.~Mohsenian-Rad are with Department of Electrical \& Computer Engineering, and M.~Raissi is with Department of Mathematics, Univ of California, Riverside. email: \{strip008,maziarr,hamedrad\}@ucr.edu.}}

\maketitle
\pagestyle{empty}
\thispagestyle{empty}

\begin{abstract}

Continuously recorded high-resolution power system waveforms capture fast dynamics and transient events. However, extracting actionable insights from these large volumes of data requires automated event detection, temporal localization, and classification. This paper proposes a spectrogram-based framework that jointly detects, localizes, and classifies events in voltage and current waveforms at the terminal of an inverter-based resource (IBR). We recast this problem as a temporal object detection problem using time-frequency representations that explicitly reveal transient and harmonic signatures. The short-time Fourier transform is applied to each waveform channel, and the resulting log-spectrograms are stacked into a multi-channel tensor and processed by a U-Net architecture to jointly estimate event occurrence, temporal boundaries, and event type. A network of the same depth and width operating on raw waveform samples serves as a baseline. Experiments on synthetic single-phase disturbances and three-phase faults in IBR terminal recordings demonstrate that the proposed approach substantially improves event detection and temporal localization compared with the raw-waveform baseline, particularly at finer time-frequency resolutions, while also improving classification accuracy.

\vspace{0.1cm}


 \end{abstract}

\section{Introduction}
\label{sec:intro}
\setlength{\parskip}{0pt}

Modern power systems are  evolving with increasing
integration of inverter-based resources (IBRs), such as solar,
wind, and battery storage, and power-electronic loads.
These resources introduce fast transients and harmonic emissions.
Typical events include switching transients, oscillations,
and incipient faults. Detecting these events can reveal
developing faults and help prevent equipment damage.
Such dynamics can be captured by Waveform Measurement
Units (WMUs) that continuously record voltage and current
waveforms at kilohertz rates~\cite{Hamed_book}.
The resulting data volumes are too large for manual
inspection, requiring automated event analysis.

A practical monitoring system must detect events,
localize their start and end times, and classify their types.
These tasks face two challenges. First, events are rare
relative to normal operation, causing class imbalance.
Second, many event signatures are small in magnitude,
making their  transient
signatures difficult to identify in raw waveform measurements.

\vspace{0.05cm}

\noindent
\textbf{Related Work:} Existing works for power system event detection can be broadly categorized into two classes: \emph{time-series based} methods
and \emph{frequency-based} methods. The time-series methods
flag an event when the signal amplitude departs
from its expected behavior. These approaches monitor
the RMS value or the peak magnitude of voltage and current, and declare an event
when a threshold is exceeded \cite{IEEE1159,Bollen_sags}. Such rules are simple and widely used in power quality monitoring. However, events whose signatures are subtle in magnitude,
such as incipient faults or low-level oscillations,
can be lost in the noise \cite{Izadi_TSG22, Izadi_TSG21}. In contrast,
frequency-based methods leverage the spectral
content of the waveform. The Fourier transform
captures harmonic and inter-harmonic components,
but discards the time information needed for localization.
Time-frequency representations recover this timing.
The short-time Fourier transform (STFT), the wavelet transform, the $S$-transform, and the Hilbert transform have been used in power-quality analysis \cite{Santoso_wavelet,Gaouda_wavelet,Dash_S,Shukla_HHT}. These
representations make transient and harmonic signatures explicit and can
capture events that remain invisible to purely amplitude-based
schemes \cite{BollenGu_book}. Recent work couples time-frequency features with learning-based classifiers, ranging from
support vector machines \cite{Axelberg_SVM} and decision trees \cite{Kumar_DT} to convolutional neural networks
 applied directly to time-frequency images \cite{Wang_CNN}.

Most existing learning-based schemes treat detection,
localization, and classification as separate stages, and many classify a short segment that is
already known to contain a disturbance. In
contrast, we address all three tasks jointly. The proposed approach
falls within the frequency-based category. Our formulation is inspired by object
detection in computer vision, where single-stage detectors such as
YOLO \cite{Redmon_YOLO} and SSD \cite{Liu_SSD} jointly predict bounding boxes, class labels, and confidence scores
\cite{Ren_FasterRCNN}. A similar approach is used in sound
detection, where events are localized in time on audio spectrograms
\cite{Mesaros_SED}. Building on these ideas, we adopt an
encoder-decoder U-Net architecture \cite{Ronneberger_UNet} that scores each temporal window for event presence and class label.

\vspace{0.05cm}

\noindent\textbf{Contributions:} The main contributions of the paper are as follows. First, we cast the joint detection, temporal localization, and classification of events in continuously recorded power systems waveform measurements as the scoring of overlapping windows of a spectrogram, followed by an explicit rule that converts the window scores into an event interval and a class. Second, we map multi-channel voltage and current measurements to a stacked log-spectrogram tensor, in which harmonic and transient signatures are explicit, and score its windows with a compact U-Net that outputs an event probability and class probabilities for each window. Third, on synthetic single-phase disturbances and on faults injected into IBR terminal recordings, with all test events taken from data that are not used for training, we compare the spectrogram input with the raw waveform under networks of the same depth and width. We show that the STFT hop governs the localization accuracy: with a fine hop, the spectrogram input detects and localizes events more accurately than the raw-waveform baseline on both tasks, whereas with the coarsest hop its event edges are less accurate.

\setlength{\parskip}{0ex plus 0.2ex minus 0.1ex}

\section{Problem Formulation}
\label{sec:problem}

We consider a WMU at the terminal of an IBR that continuously records voltage and current waveforms at the sampling rate $f_s$. The recorded stream is divided into records of $N$ samples. A record is the sequence $x[n]\in\mathbb{R}^C$, $n=0,\ldots,N-1$, where $C$ is the number of channels and $x_c[n]$ is channel~$c$. We consider a single-phase voltage, with $C=1$, and the three phase voltages and currents, with $C=6$ and $x[n]=\big[v_a[n],v_b[n],v_c[n],i_a[n],i_b[n],i_c[n]\big]^{\top}$.

An \emph{event} is an interval in which the waveform departs from its normal sinusoidal behavior, e.g., due to harmonic distortion or a short-circuit fault. Each record contains one event, which starts at sample $s$, ends at sample $e$, and has a class $y\in\mathcal{Y}$, e.g., the fault type. Given a record, a detector must decide whether it contains an event (\emph{detection}), when the event starts and ends (\emph{localization}), and what type of event it is (\emph{classification}). Thus, it returns either ``no event'' or an estimated interval $[\hat s, \hat e]$, a class $\hat y$, and a confidence $\hat p\in[0,1]$. Localization is measured by the intersection over union
\begin{align}\label{eq:iou}
    \text{IoU}=\frac{\big|[\hat s, \hat e]\cap [s, e]\big|}{\big|[\hat s, \hat e]\cup [s, e]\big|},
\end{align}
where $|\cdot|$ is the length of an interval. For example, an estimate that covers the event but is 25\% longer has $\text{IoU}=0.8$. A detection is correct at threshold $\alpha$ if $\text{IoU}\geq \alpha$.

\section{Methodology}

\label{sec:spectrogram}

\begin{figure*}[t]
  \centering
  \begin{tikzpicture}
    \node[anchor=south west, inner sep=0] (img) at (0,0)
      {\includegraphics[width=\textwidth]{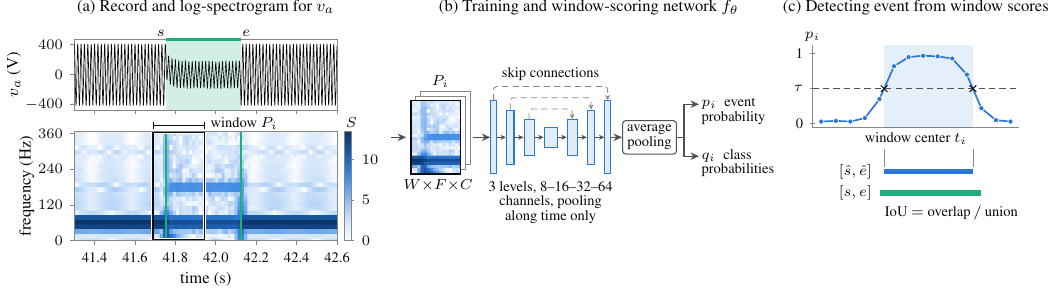}};
    \begin{scope}[x={(img.south east)}, y={(img.north west)}]
      \node[anchor=south west, inner sep=2pt, rounded corners=2pt,
            draw=black!30, fill=black!3, font=\scriptsize, align=left] at (0.405, 0.012)
      {\textbf{Pipeline:} $x_c[n]\xrightarrow{\text{STFT, \eqref{eq:stft}--\eqref{eq:tensor}}}\mathbf{Z}\xrightarrow{\text{windows, \eqref{eq:window}}}P_i\xrightarrow{\text{U-Net, \eqref{eq:net}}}(p_i,q_i)\xrightarrow{\text{rule, \eqref{eq:edges}}}[\hat s,\hat e],\,\hat y,\,\hat p$\\[0.5pt]
       \textbf{Symbols:} $L$ frame length, $H$ hop, $F$ bins, $W$ frames per window, $\Delta$ stride, $\tau$ threshold};
    \end{scope}
  \end{tikzpicture}\par\vspace{-0.2cm}
  \caption{Overview of the detector. (a) Faulted-phase voltage of a three-phase test record with a single-line-to-ground fault on $[s,e]$, and its log-spectrogram \eqref{eq:logspec}.
  (b) Every window $P_i$ is mapped to an event probability $p_i$ and to class probabilities $q_i$. (c) An event is detected when $p_i\geq \tau$.}
  \label{fig:overview}
  \vspace{-0.25cm}
\end{figure*}

Fig.~\ref{fig:overview} illustrates the four steps of the proposed detector, which are described in Sections~\ref{sec:logspec}--\ref{sec:decoding}.

\subsection{Log-Spectrogram Representation}
\label{sec:logspec}
The STFT computes the spectrum of short segments of the waveform, called frames. Each channel is divided into frames of $L$ samples, where frame $m$ begins at sample $mH$ and the hop $H\le L$ is the shift between consecutive frames; e.g., $H=L/4$ gives a 75\% overlap. This gives $M=\lfloor (N-L)/H\rfloor+1$ frames. Frame $m$ is tapered by a Hamming window $w[n]$, which limits the leakage of the strong fundamental into the neighboring bins, and is transformed by an $L$-point discrete Fourier transform,
\begin{equation}
  X_c[m,k]=\sum_{n=0}^{L-1} w[n]\,x_c[mH+n]\,e^{-j2\pi kn/L},
  \label{eq:stft}
\end{equation}
where bin $k$ corresponds to the frequency $kf_s/L$. Thus, the bins are $f_s/L$ apart in frequency, and the frames are $H/f_s$ apart in time. The log-spectrogram keeps the $F$ lowest bins, $k=0,\dots,F-1$:
\begin{equation}
  S_c[m,k]=\log\!\Big(1+\tfrac{1}{L}\big|X_c[m,k]\big|^{2}\Big).
  \label{eq:logspec}
\end{equation}
The retained band covers the fundamental and the low-order harmonics, while the higher bins carry mostly noise. The logarithm compresses the range between the fundamental and the much weaker harmonic and transient components, so that both remain visible. Stacking the channels yields the tensor
\begin{equation}
  \mathbf{Z}\in\mathbb{R}^{M\times F\times C},\qquad \mathbf{Z}[m,k,c]=S_c[m,k],
  \label{eq:tensor}
\end{equation}
whose axes are time (frames), frequency, and channel. Fig.~\ref{fig:overview}(a) shows one channel of $\mathbf{Z}$: the fundamental forms a dark horizontal band, and the fault adds a band near 180~Hz and broadband energy at its inception and clearing.

\subsection{Window Scoring}
\label{sec:windows}
To examine the record piece by piece, a window of $W$ consecutive frames slides along the time axis of $\mathbf{Z}$ in steps of $\Delta$ frames (the \emph{stride}). Window $i$ contains frames $i\Delta$ to $i\Delta+W-1$ of all bins and channels,
\begin{equation}
  P_i=\mathbf{Z}[\,i\Delta:i\Delta+W-1,\;:\,,\;:\,]\in\mathbb{R}^{W\times F\times C},
  \label{eq:window}
\end{equation}
for $i=0,\dots,I-1$ with $I=\lfloor (M-W)/\Delta\rfloor+1$. It covers the $N_w=(W-1)H+L$ samples that begin at sample $i\Delta H$, so its center is at sample
\begin{equation}
  t_i=i\Delta H+N_w/2 .
  \label{eq:center}
\end{equation}
For training, window $i$ receives the label $z_i=1$ if its center lies inside the event, $s\le t_i\le e$, and $z_i=0$ otherwise.

\subsection{Training}
\label{sec:training}
A network $f_\theta$ maps every window to an event probability $p_i$ and to class probabilities $q_i$, where $q_i(y)$ is the probability of class $y$:
\begin{equation}
  (p_i,\,q_i)=f_\theta(P_i),\qquad p_i\in[0,1],\quad q_i\in[0,1]^{|\mathcal{Y}|}.
  \label{eq:net}
\end{equation}
We use a small U-Net \cite{Ronneberger_UNet}, shown in Fig.~\ref{fig:overview}(b). Its encoder repeatedly convolves and downsamples the input, so that deeper levels respond to longer patterns, e.g., a sustained harmonic rather than a short transient; its decoder upsamples the result, and skip connections pass the fine-grained encoder features to the decoder. It has three encoder levels, a bottleneck, and three decoder levels. Each consists of one $3\times3$ convolution followed by a ReLU, and the encoder levels and the bottleneck have 8, 16, 32, and 64 feature maps. Pooling and upsampling act along time only, so the frequency resolution, which separates, e.g., a harmonic from an interharmonic, is preserved. The output feature map is averaged over time and frequency and passed to a linear layer with a softmax to give $p_i$; a second linear layer with a softmax, applied to the feature map averaged over time only, gives $q_i$. The network has about 62{,}000 parameters. It is trained by minimizing
\begin{equation}
  \mathcal{L}(\theta)=\mathcal{L}_{\mathrm{detection}}+\lambda\,\mathcal{L}_{\mathrm{classification}},
  \label{eq:loss}
\end{equation}
where $\lambda>0$ weighs the two tasks, $\mathcal{L}_{\mathrm{detection}}$ is the cross-entropy between $p_i$ and $z_i$ over all training windows, and $\mathcal{L}_{\mathrm{classification}}$ is the cross-entropy between $q_i$ and the class $y$ over the event windows only. Since most windows contain no event, an unweighted $\mathcal{L}_{\mathrm{detection}}$ would favor always predicting ``no event''; hence, the event windows are weighted by the ratio of non-event to event windows.

\subsection{Detecting Event From Window Scores}
\label{sec:decoding}
At inference, the windows of a record give the confidence trace $(t_i,p_i)$, $i=0,\dots,I-1$, sketched in Fig.~\ref{fig:overview}(c), which rises toward 1 inside the event. Given a threshold $\tau$, an event is declared if at least $n_{\min}$ consecutive windows satisfy $p_i\ge\tau$, which suppresses isolated false alarms. Among such runs, the longest one, $i_1,\dots,i_2$, is retained, and its edges are placed where the linearly interpolated trace crosses $\tau$:
\begin{equation}
\begin{aligned}
  \hat s&=t_{i_1-1}+\frac{\tau-p_{i_1-1}}{p_{i_1}-p_{i_1-1}}\,\big(t_{i_1}-t_{i_1-1}\big),\\
  \hat e&=t_{i_2}+\frac{p_{i_2}-\tau}{p_{i_2}-p_{i_2+1}}\,\big(t_{i_2+1}-t_{i_2}\big),
\end{aligned}
\label{eq:edges}
\end{equation}
with $\hat s=t_{i_1}$ if $i_1=0$ and $\hat e=t_{i_2}$ if $i_2=I-1$. The class is taken from the most confident window, $\hat
y=\arg\max_{y} q_{i^\star}(y)$ with
$i^\star=\arg\max_i p_i$, and the confidence of the detection is $\hat p=\max_i p_i$. If no such run exists, the record is declared event-free. As the trace is sampled every $\Delta H$ samples, the hop $H$ sets both the time span $N_w$ of a window and the resolution of the edges. The threshold $\tau$ is the value that maximizes the window-level F1 score on the validation records.

\subsection{Raw-Waveform Baseline Network (U-Net-1D)}
\label{sec:baseline}
To isolate the effect of the input, the baseline replaces $P_i$ by the raw samples of the same time span, $U_i\in\mathbb{R}^{N_w\times C}$, and the $3\times3$ convolutions by one-dimensional convolutions of length~5. Depth, widths, pooling, loss, and decoding are unchanged, and the network has about 34{,}000 parameters. Its class head reads the time-averaged feature map and is trained without updating the shared layers. We refer to the spectrogram network as U-Net-2D.

\subsection{Evaluation Metrics}
\label{sec:metrics}
We evaluate detection, localization, and classification separately:

\paragraph{Localization} It is measured by the mean IoU \eqref{eq:iou} over the declared detections.

\paragraph{Detection} At a fixed IoU threshold $\alpha$, a declared
interval is a true positive (TP) if $\text{IoU}\ge\alpha$; any other declared interval is a false positive (FP), i.e., a false alarm, and an event without a TP is a false negative (FN), i.e., a missed event. The precision and recall are
\begin{equation}
\mathrm{Prec}=\frac{\mathrm{TP}}{\mathrm{TP}+\mathrm{FP}},\qquad
\mathrm{Rec}=\frac{\mathrm{TP}}{\mathrm{TP}+\mathrm{FN}}.
\label{eq:pr}
\end{equation}
Sweeping a cut-off on the confidence $\hat p$ at a fixed $\alpha$ traces a
precision-recall curve with points $(\mathrm{Rec}_n,\mathrm{Prec}_n)$, sorted by recall. The average precision is the area under its interpolated version,
\begin{align}
\begin{aligned}
&\mathrm{AP}=\sum_{n}\big(\mathrm{Rec}_{n+1}-\mathrm{Rec}_n\big)\,\mathrm{Prec}_{\mathrm{int}}(\mathrm{Rec}_{n+1}), \\
&\mathrm{Prec}_{\mathrm{int}}(r)=\max_{\tilde r\ge r}\mathrm{Prec}(\tilde r),
\label{eq:ap}
\end{aligned}
\end{align}
which equals 1 only if all events are found without false alarms. AP@$\alpha$ is the AP at the IoU threshold $\alpha$, computed over all test events regardless of class, and
AP$_{[.5:.95]}$ is the mean of AP@$\alpha$ over $\alpha\in\{0.50, 0.55, \dots, 0.95\}$, which rewards accurate localization.

\paragraph{Classification} The class accuracy is the fraction of the true-positive detections at $\alpha$ that are assigned the correct class.

\providecommand{\placeholder}[1][--]{\textcolor{red}{\textbf{#1}}}

\begin{table*}[t]
\centering
\caption{Results on the 40 single-phase test disturbances}
\label{tab:sp_results}
\footnotesize
\begin{tabular*}{\textwidth}{@{\extracolsep{\fill}}ll cccc c c@{}}
\toprule
 & & \multicolumn{4}{c}{Detection} & Localization & Classification\\
\cmidrule(lr){3-6}\cmidrule(lr){7-7}\cmidrule(l){8-8}
Detector & Overlap ($H$) & AP@0.5 & AP@0.7 & AP$_{[.5:.95]}$ & Recall & IoU & Class accuracy\\
\midrule
U-Net-1D (raw waveform) & ---       & 0.650 & 0.550 & 0.454 & 0.650 & 0.831 & 0.654\\
\midrule
U-Net-2D (spectrogram)  & 0\% (64)  & 0.949 & 0.705 & 0.557 & 0.950 & 0.796 & 0.974\\
                        & 25\% (48) & 0.975 & 0.891 & 0.723 & 0.975 & 0.855 & 0.974\\
                        & 50\% (32) & \textbf{1.000} & \textbf{1.000} & 0.813 & \textbf{1.000} & 0.898 & \textbf{0.975}\\
                        & 75\% (16) & \textbf{1.000} & 0.975 & \textbf{0.904} & \textbf{1.000} & \textbf{0.941} & \textbf{0.975}\\
\bottomrule
\end{tabular*}\par\vspace{-0.15cm}
\end{table*}

\section{Experiments}
\label{sec:experiments}
We now evaluate the proposed framework on two tasks at the terminal of an IBR: single-phase disturbance detection and three-phase fault detection. In both, the window stride is $\Delta=8$ frames in training and 4 in testing, $\lambda=0.5$ for U-Net-2D and 1 for U-Net-1D, and $\tau$ is selected on the validation records. Tables~\ref{tab:sp_results} and~\ref{tab:tp_results} report AP@0.5, AP@0.7, AP$_{[.5:.95]}$, and the recall at $\alpha=0.5$ for detection, the mean IoU for localization, and the class accuracy. The open source dataset and code repository are available at: \mbox{\href{https://github.com/shivanshutripath/WMU-Spectrogram}{https://github.com/shivanshutripath/WMU-Spectrogram}}.

\subsection{Dataset}
We use two datasets for single-phase and three-phase fault detection.

\begin{figure}[t]
  \centering
  \includegraphics[width=\columnwidth]{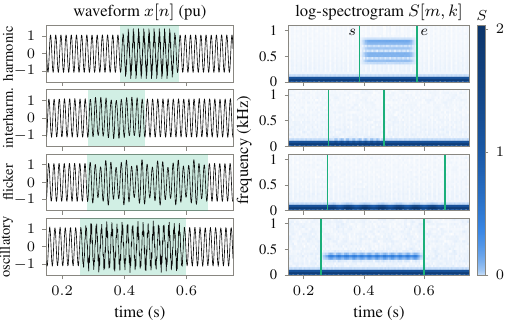}\par\vspace{-0.2cm}
  \caption{One test record of each single-phase disturbance class: waveform with the event $[s,e]$ shaded (left) and log-spectrogram \eqref{eq:logspec} at 75\% overlap (right; color scale compressed by the power $0.3$).}
  \label{fig:overview1}
  \vspace{-0.25cm}
\end{figure}

\emph{Single-phase disturbances:} Each record is a 1-s voltage waveform ($C=1$) sampled at $f_s=3$~kHz ($N=3000$): a 60-Hz, 1-pu sinusoid with zero-mean Gaussian noise and one disturbance on $[s, e]$, namely a harmonic distortion, an interharmonic distortion, a voltage flicker, or an oscillatory transient. The dataset has six sets of 40 records, balanced across the classes, i.e., 240 records and 720{,}000 samples. Five sets are split at random into 170 training and 30 validation records, and the sixth set provides the 40 test records. Fig.~\ref{fig:overview1} shows one test record of each class.

\emph{Three-phase faults:} Each record is a 60-s recording of the three phase voltages and currents at an IBR terminal ($C=6$), sampled at $f_s=1$~kHz ($N=60{,}000$), with one injected single-line-to-ground, line-to-line, double-line-to-ground, or three-phase fault. The 200 records contain $200\times6\times60{,}000=72$ million samples. Each record is tagged with one of three recording backgrounds, $A$, $B$, and $C$, that differ in noise conditions. We use 100 records for training and 34 for validation, all from backgrounds~$A$ and~$B$, and the 66 records of the unseen background~$C$ for testing, so the results also measure generalization across noise. Fig.~\ref{fig:overview}(a) shows part of a test record.

\subsection{Single-Phase Disturbance Detection}
\label{sec:sp}
U-Net-2D uses a Hamming window of $L=64$ samples (21~ms, with bins 46.9~Hz apart), the $F=24$ lowest bins (0--1078~Hz), and the hops $H=64$, 48, 32, and 16 samples, i.e., overlaps of 0, 25, 50, and 75\% ($M=46$, 62, 92, and 184 frames). Each window has $W=16$ frames, and an event requires $n_{\min}=2$ consecutive windows. U-Net-1D uses the window grid of the 75\% setting. Each class has a distinct signature in the log-spectrogram (Fig.~\ref{fig:overview1}): harmonics add bands at odd multiples of 60~Hz, the interharmonic adds a band between them, flicker modulates the fundamental band, and the oscillatory transient adds a band at its frequency.

\begin{figure}[t]
  \centering
  \includegraphics[width=\columnwidth]{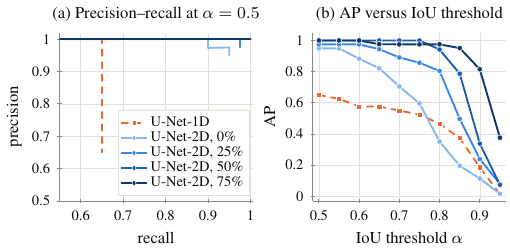}\par\vspace{-0.2cm}
  \caption{Detection performance on the 40 single-phase test events: (a) precision-recall curves at $\alpha=0.5$; (b) AP versus the IoU threshold $\alpha$.}
  \label{fig:sp_curves}
\end{figure}

Table~\ref{tab:sp_results} and Fig.~\ref{fig:sp_curves} show the results; Fig.~\ref{fig:sp_curves}(b) shows how the AP drops as the required IoU $\alpha$ increases. U-Net-2D detects 38--40 of the 40 test events with IoU${}\ge0.5$, against 26 for U-Net-1D, keeps a precision of 1 up to a recall of at least 0.90, against 0.65 (Fig.~\ref{fig:sp_curves}(a)), and assigns the correct class to 97\% of its correct detections at every overlap, against 65\%. Localization depends on the hop. At 0\% overlap, a window (0.34~s) is longer than 29 of the 40 test events, and the mean IoU of U-Net-2D (0.796) is below that of U-Net-1D (0.831); from 25\% overlap on it is higher, reaching 0.941 at 75\%. AP$_{[.5:.95]}$ rises from 0.557 to 0.904 with the overlap, against 0.454 for U-Net-1D, and at $\alpha=0.9$ only the 75\% setting keeps an AP above 0.8 (Fig.~\ref{fig:sp_curves}(b)).

\subsection{Three-Phase Fault Detection}
\label{sec:tp}
The spectrogram uses the same $L$, $F$, $W$, and overlaps. At $f_s=1$~kHz, the bins are 15.6~Hz apart and cover 0--359~Hz. The overlaps give $M=937$, 1249, 1874, and 3747 frames, windows of $N_w=1024$, 784, 544, and 304 samples, and $\mathbf{Z}\in\mathbb{R}^{M\times 24\times 6}$; an event requires $n_{\min}=3$ consecutive windows. A fault changes the voltages and currents of the faulted phases at once, and the stacked tensor lets the network compare these changes across the six channels.

\begin{figure}[t]
  \centering
  \includegraphics[width=\columnwidth]{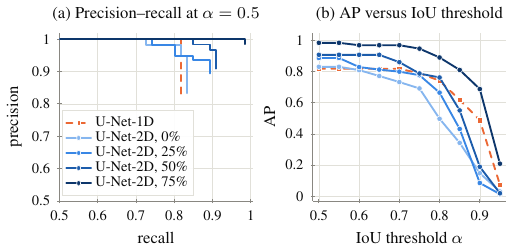}\par\vspace{-0.2cm}
  \caption{Detection performance on the 66 three-phase test faults: (a) precision-recall curves at $\alpha=0.5$; (b) AP versus the IoU threshold $\alpha$.}
  \label{fig:tp_curves}
  \vspace{-0.25cm}
\end{figure}

Table~\ref{tab:tp_results} and Fig.~\ref{fig:tp_curves} show the results. At 75\% overlap, U-Net-2D is ahead of U-Net-1D in every column: it detects 65 of the 66 faults (recall 0.98, against 0.82; Fig.~\ref{fig:tp_curves}(a)), raises AP@0.5 from 0.818 to 0.985 and AP$_{[.5:.95]}$ from 0.680 to 0.842, and reaches a mean IoU of 0.926, against 0.902. At coarser hops, its AP@0.5 is still higher (0.831--0.908), but its windows (0.544--1.024~s) are longer than many test faults, and its mean IoU (0.839--0.853) falls below that of U-Net-1D. Accordingly, at $\alpha=0.9$ the AP is 0.688 at 75\% overlap, 0.49 for U-Net-1D, and below 0.19 at the coarser hops (Fig.~\ref{fig:tp_curves}(b)). The class accuracy is 0.32--0.39 for U-Net-2D and 0.30 for U-Net-1D, close to the 0.30 obtained by always predicting the most frequent test class, likely because the test faults involve phase combinations that are rare or absent in training. Both test sets are small (one event changes the recall by 0.025 or 0.015), and each network was trained once.

\begin{table*}[t]
\centering
\caption{Results on the 66 three-phase test faults of the unseen background~$C$}
\label{tab:tp_results}
\footnotesize
\begin{tabular*}{\textwidth}{@{\extracolsep{\fill}}ll cccc c c@{}}
\toprule
 & & \multicolumn{4}{c}{Detection} & Localization & Classification\\
\cmidrule(lr){3-6}\cmidrule(lr){7-7}\cmidrule(l){8-8}
Detector & Overlap ($H$) & AP@0.5 & AP@0.7 & AP$_{[.5:.95]}$ & Recall & IoU & Class accuracy\\
\midrule
U-Net-1D (raw waveform) & ---       & 0.818 & 0.818 & 0.680 & 0.82 & 0.902 & 0.30\\
\midrule
U-Net-2D (spectrogram)  & 0\% (64)  & 0.831 & 0.733 & 0.569 & 0.83 & 0.839 & 0.36\\
                        & 25\% (48) & 0.888 & 0.800 & 0.619 & 0.89 & 0.853 & \textbf{0.39}\\
                        & 50\% (32) & 0.908 & 0.861 & 0.680 & 0.91 & 0.844 & 0.32\\
                        & 75\% (16) & \textbf{0.985} & \textbf{0.969} & \textbf{0.842} & \textbf{0.98} & \textbf{0.926} & 0.38\\
\bottomrule
\end{tabular*}\par\vspace{-0.15cm}
\end{table*}

\section{Conclusions and Future Work}\label{sec:conclusion}
This paper presented a spectrogram-based framework for the joint detection, localization, and classification of power system events from continuous waveform measurements at the terminal of an IBR. In experiments, the spectrogram input detected more events than a raw-waveform network of the same size on both tasks, including faults recorded under unseen noise conditions. It also classified the single-phase disturbances far more accurately and, with a fine STFT hop, localized the events more accurately. Future work can extend the framework in several directions like online detection, which scores the windows in real time as the samples stream in and using time-frequency representations, such as wavelets on our framework.

\bibliographystyle{IEEEtran}
\bibliography{Ref}

\end{document}